\documentclass[journal]{IEEEtran}
\IEEEoverridecommandlockouts
\usepackage{algorithmic}
\usepackage{array}
\usepackage{cite}
\usepackage{amsmath,amssymb,amsfonts}
\usepackage{algorithmic}
\usepackage{graphicx}
\usepackage{textcomp}
\usepackage{xcolor}
\usepackage{caption}
\usepackage{subcaption}
\def\BibTeX{{\rm B\kern-.05em{\sc i\kern-.025em b}\kern-.08em
    T\kern-.1667em\lower.7ex\hbox{E}\kern-.125emX}}

\begin{document}
\title{Analog Compressed Sensing for Sparse Frequency Shift Keying Modulation Schemes}

\author{Kathleen Yang,~\IEEEmembership{Student member, IEEE,}
        Diana C. Gonz\'{a}lez,~\IEEEmembership{}
        Yonina C. Eldar,~\IEEEmembership{Fellow, IEEE,}
        Muriel M\'{e}dard,~\IEEEmembership{Fellow,~IEEE}
        % <-this % stops a space
\thanks{
Kathleen Yang and Muriel Médard are with
the Research Laboratory of Electronics, MIT, Cambridge, MA 02139 USA
(e-mail: klyang@mit.edu; medard@mit.edu).
}% <-this % stops a space
\thanks{
Diana C. Gonz\'{a}lez was affiliated with MIT, Cambridge, MA 02139 USA
(e-mail: dianigon@decom.fee.unicamp.br).
}% <-this % stops a space
\thanks{
Yonina Eldar is affiliated with Weizmann Institute of Science, Rehovot, Israel
(e-mail: yonina.eldar@weizmann.ac.il).
}% <-this % stops a space
\thanks{This work was supported by the Office of Naval Research under the Battelle prime contract N6833518C0179.}% <-this % stops a space
\thanks{Manuscript received ; revised.}
\thanks{Part of this work has been accepted in SPAWC 2022 \cite{ifsk_cs_conf}.
We expand on our previous work by exploring the performance of the compressed sensing receiver when used to receive wideband time frequency coding signals. We also incorporate a discussion on the differences in hardware costs and complexity between the compressed sensing receiver and the bank of frequency-selective filters.}}

\maketitle

\begin{abstract}
There is a growing interest in signaling schemes that operate in the wideband regime due to the crowded frequency spectrum.
However, a downside of the wideband regime is that obtaining channel state information is costly, and the capacity of previously used modulation schemes such as code division multiple access and orthogonal frequency division multiplexing begins to diverge from the capacity bound without channel state information.
Impulsive frequency shift keying and wideband time frequency coding have been shown to perform well in the wideband regime without channel state information, thus avoiding the costs and challenges associated with obtaining channel state information.
However, the maximum likelihood receiver is a bank of frequency-selective filters, which is very costly to implement due to the large number of filters.
In this work, we aim to simplify the receiver by using an analog compressed sensing receiver with chipping sequences as correlating signals to detect the sparse signals.
Our results show that using a compressed sensing receiver allows for the simplification of the analog receiver with the trade off of a slight degradation in recovery performance.
%Our results show that the performance of the bank of frequency-selective filters is consistently 1.4-1.5 times that of the compressed sensing receiver when the number of correlating signals is equivalent to the number of frequency-selective filters for both impulsive frequency shift keying and wideband time frequency coding signals.
For a fixed frequency separation, symbol time, and peak SNR, the performance loss remains the same for a fixed ratio of number of correlating signals to the number of frequencies.
\end{abstract}


% \fi
\IEEEpeerreviewmaketitle

\section{Introduction}
\IEEEPARstart{T}{here} is increasing research on the performance of signaling schemes in the wideband regime, with larger bandwidths at higher frequencies, in order to achieve higher data rates and to avoid the crowded frequency spectrum below 3 GHz \cite{mmwave_rappaport}.
The noise and fading associated with the wideband regime introduces new challenges to previously used signaling schemes such as orthogonal frequency division multiplexing (OFDM), which was used in 4G and LTE networks, and code division multiple access (CDMA), which was used in 3G networks.
Both of these signaling schemes require channel state information (CSI) in order to achieve reliable and high data rates. 
However, the bandwidth, fading, and noise associated with the wideband regime makes it challenging and costly to obtain CSI in this regime \cite{6G_non_coherent}.

The capacities of OFDM and CDMA without CSI begin to fail and diverge from the capacity bound in the wideband regime.
The capacity of CDMA is inversely proportional to the bandwidth in the wideband regime due to a fourth moment constraint \cite{cdma_wideband} and insufficient fourthegy per unit energy to achieve reliable bits \cite{fourthegy}.
The capacity of OFDM with phase shift keying (PSK) was shown to decrease once the bandwidth of the system exceeds the critical bandwidth at which OFDM's capacity reaches a maximum \cite{ofdm_wideband}.

Therefore, there is great interest in exploring signaling schemes that can perform well in the wideband regime without CSI, especially for 6G and beyond communications \cite{6G_non_coherent}.
Kennedy showed that signals concentrated in both time and frequency perform well in highly doubly dispersive channels, which are associated with the wideband regime \cite{kennedy_fading}.
In particular, impulsive frequency shift keying (I-FSK) and wideband time frequency coding (WTFC) have been shown to perform well in this regime without CSI \cite{teletar_pfsk, multitone_pfsk, wtfc, milcom}.
I-FSK incorporates FSK with a duty cycle, which results in a signal that is impulsive in both time and frequency.
The amplitude of the I-FSK signal is multiplied by the inverse of the duty cycle, and the signal itself is transmitted once per cycle \cite{teletar_pfsk}.
As the bandwidth of the system approaches infinity, and the duty cycle approaches zero, I-FSK achieves the capacity of multipath fading channels \cite{teletar_pfsk}.
In finite bandwidths and non-zero duty cycles. I-FSK can achieve rates close to the energy-limited capacity bound of multipath fading channels \cite{multitone_pfsk}.
%Over-the-air testing of I-FSK was performed, and showed the feasibility of I-FSK \cite{milcom}.
WTFC builds on top of I-FSK and incorporates pulse position modulation (PPM) with I-FSK, thus allowing for information to be encoded in the time period that the signal is transmitted.
Under the same channel conditions, symbol time, and duty cycle, WTFC's capacity is greater than I-FSK's with the trade off of having a greater symbol error rate \cite{wtfc}.

In general, I-FSK and WTFC perform well in the wideband regime without CSI, but a drawback of these schemes is that the maximum likelihood receiver is a bank of frequency-selective filters, which are otherwise known as matched filters for FSK signals \cite{proakis}.
A bank of frequency-selective filters is costly and difficult to implement due to the large number of filters required, which increases linearly with the bandwidth of the system.
In addition, the power consumption and power dissipation associated with this analog receiver often renders them impractical to implement \cite{analog_vs_digital_circuits, low_power_design}.
The power consumption of the bank of frequency-selective filters is proportional to the noise level and the bandwidth, which can grow to be very large in the wideband regime \cite{analog_power_consumption}.
Instead of using a bank of frequency-selective filters, digital signal processing can be used to detect I-FSK and WTFC signals, as done in \cite{milcom} for I-FSK signals.
Analog-to-digital converters (ADCs) are needed for digital processing, but high rate ADCs are expensive and may have high power consumption depending on the desired bit resolution, which increases the cost of digital signal processing \cite{CS_duarte, adc_power, adc_parameters}.
%Thus, there is a need to investigate receivers that can reduce these costs and difficulties. 

%Fast Fourier transforms (FFT) can be used in digital systems to detect I-FSK and WTFC signals as done in \cite{milcom} for I-FSK signals, but FFTs also become costly as the cost of analog-to-digital converters (ADC) increase with increasing bandwidths.

Instead of using a bank of frequency-selective filters or digital signal processing to recover I-FSK and WTFC signals, we follow prior work on analog compressed sensing receivers \cite{yonina_cs_analog, yonina_textbook, rdmud, pscs, aic_rd, random_demodulator, efficient_rand_demod, rand_demod_stat, rand_demod_cons, mult_chip, mwc} and suggest using a reduced-rate receiver based on correlating signals.
In particular, we use pseudo-noise correlating signals/chipping sequences, which can be implemented with cascaded shift registers \cite{sklar_digital_comms, cdma_handbook}. 
%and are less costly than a bank of frequency-selective filters and large bandwidth ADCs \cite{sklar_digital_comms, cdma_handbook}.
%Despite its poor performance in the wideband regime without CSI, CDMA has an advantage over I-FSK and WTFC due to its receiver using a bank of pseudo-noise correlating signals for demodulation \cite{sklar_digital_comms}.
%Implementing pseudo-noise correlating signals using cascaded shift registers is less costly than implementing a bank of frequency-selective filters or using ADCs that can sample a large bandwidth for FFTs \cite{cdma_handbook}.
%This leads to a desire to investigate the performance of a bank of correlating signals when recovering I-FSK and WTFC signals.
%Using correlating signals in the receiver is one method of simplifying the receiver for I-FSK and WTFC signals, and other techniques can also be used to further reduce its complexity.
The sparsity of I-FSK and WTFC signals, where the number of transmitted frequencies is much less than the number of available frequencies, enables the usage of sub-Nyquist receivers with a set of correlating signals smaller in size than the set of basis signals \cite{yonina_cs_analog, yonina_textbook}.
This approach can reduce the amount of hardware required in the receiver, which makes the receiver more practical compared to a bank of frequency-selective filters and digital signal processing.
Recovery algorithms, such as orthogonal matching pursuit (OMP) and thresholding, are then used to recover the noisy and faded signals \cite{yonina_textbook, omp, noisy_signal_recovery}.
%While the recovery algorithms are more complex than a square-law comparison used for the outputs of a bank of frequency-selective filters, the algorithm is performed digitally, and thus does not introduce complexity in the hardware.

The performance of the recovery algorithms using correlating signals to recover analog signals such as FSK signals have previously been investigated, though primarily in additive white Gaussian noise (AWGN) channels or in scenarios where the receiver knows the CSI.
The reduced dimension multi-user detection receiver used a bank of correlating receivers, which are generated from a linear combination of the users' signals, to detect active users signals that passed through an AWGN channel with CSI known at the receiver \cite{rdmud}.
The parallel segmented compressed sensing structure contained a bank of filters where overlapping portions of the noisy signal were sampled in parallel and reconstructed using the OMP algorithm \cite{pscs}.
The random demodulator used a chipping sequence, which are signals that are composed of rectangular pulses of length less than or equal to the Nyquist rate, to recover noisy and noiseless analog signals in band-limited systems by sampling the output multiple times with an ADC \cite{aic_rd, random_demodulator, efficient_rand_demod, rand_demod_stat, rand_demod_cons}.
The multichannel random demodulator, which contains a bank of chipping sequences, was used to recover noisy OFDM signals, which reduced the ADC requirements but increased the overall amount of hardware \cite{mult_chip}.
The modulated wideband converter also used multiple periodic chipping sequences to reconstruct noisy multiband analog signals \cite{mwc}.

%Due to chipping sequences being less costly and more practical than FFTs and a bank of frequency-selective filters in large bandwidth systems, there is interest in investigating how they perform when recovering I-FSK and WTFC signals.
In this work, we investigate the performance of a compressed sensing receiver with chipping sequences as the correlating signals and an integrator as a low-pass filter to recover I-FSK and WTFC signals. 
We recover FSK-based analog signals that have undergone Rayleigh fading and AWGN, without CSI at the receiver.
In contrast, previous works have primarily focused on the cases with only AWGN or AWGN with CSI.
We compare the performance of a chipping sequence-based compressed sensing receiver to a bank of frequency-selective filters when using OMP to recover I-FSK and thresholding to recover WTFC signals with AWGN and multipath fading.
We demonstrate that the compressed sensing receiver performs comparably to the frequency-selective filters with an equivalent number of correlating signals, and explore the trade off between the number of correlating signals and the probability of a symbol error.
We demonstrate that under certain conditions such as a fixed frequency separation, symbol time, and peak SNR, the ratio of the recovery using the compressed sensing receiver to that of the bank of frequency-selective receivers remains constant, which may allow for extrapolation to larger bandwidth systems to evaluate the trade offs when using a compressed sensing receiver versus a frequency-selective receiver in order to save computation time.

This paper is organized as follows. 
In Section \ref{sec:signal_model}, we discuss the I-FSK and WTFC signal model inputs and the Rayleigh fading channel outputs. 
In Section \ref{sec:receivers}, we consider receivers with a bank of frequency-selective filters and receivers with a bank of chipping sequences and derive the outputs of these receivers with I-FSK and WTFC signals.
We then compare the performance of the bank of frequency-selective filters versus the performance of the compressed sensing receiver with a bank of chipping sequences when recovering I-FSK and WTFC signals, and discuss the hardware complexity and costs for both receivers in Section \ref{sec:results}.
We end with concluding remarks in Section \ref{sec:conclusion}.

%Part of this work has been previously published in \cite{IFSK_cs}.
%This work builds upon the prior work by including the recovery of WTFC signals as well as introducing more simulation results at large bandwidths due to the interest of using the I-FSK and WTFC signaling schemes in the wideband regime.

Throughout this paper we use the following notations: 
The complex conjugate of an analog signal is denoted with an over-bar. 
The absolute value is written as $|\cdot|$.
Matrices and vectors are represented by bold capital and lower case letters or symbols respectively, e.g. $\mathbf{A}$ and $\mathbf{a}$. 
The conjugate transpose of a matrix or vector is denoted by $*$, e.g. $\mathbf{A}^*$. 
%In the case of a real matrix or vector, the conjugate transpose is simply the transpose. 
The expectation of a random variable is written as $\mathbb{E}[\cdot]$.

\section{Signal Model}
\label{sec:signal_model}

\begin{figure*}[t]
\centering
\begin{subfigure}{0.49\textwidth}
    \centering
    \includegraphics[width=\textwidth]{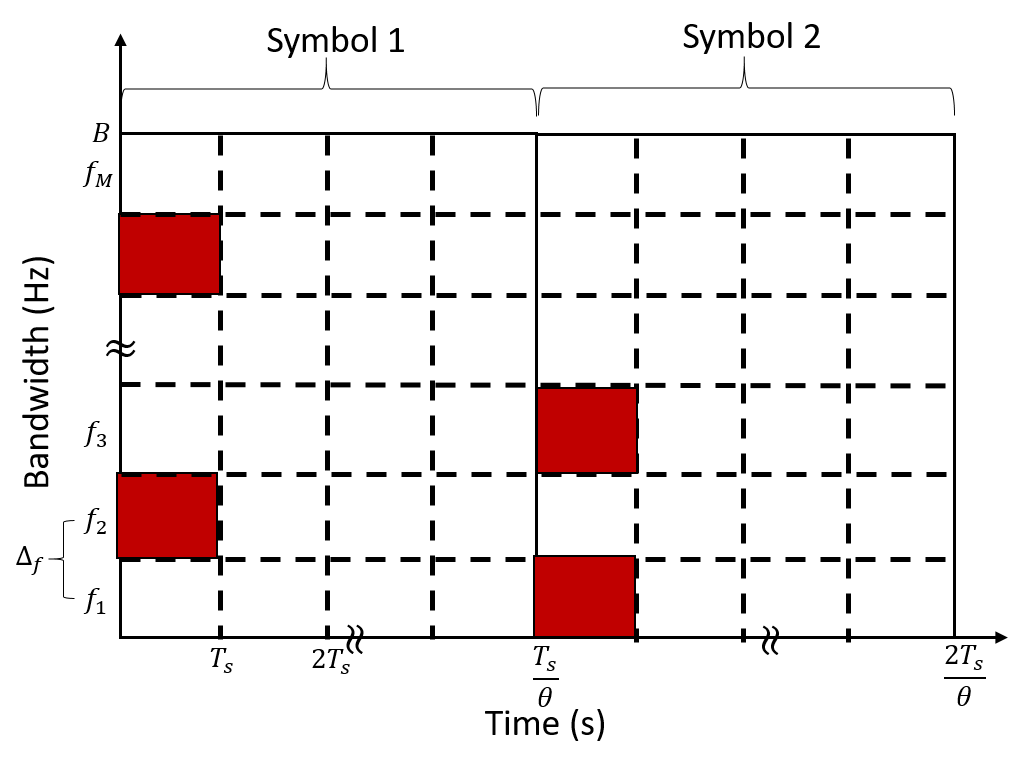}
    \caption{Grid illustrating the transmission of two I-FSK symbols with $Q=2$ frequencies randomly chosen from $M$ possible frequencies and duty cycle of $\theta$. Shaded regions denote transmitted frequencies. I-FSK signals are restricted to a known time period.}
    \label{fig:PFSK_grid}
\end{subfigure}
\hfill
\begin{subfigure}{0.49\textwidth}
    \centering
    \includegraphics[width=\textwidth]{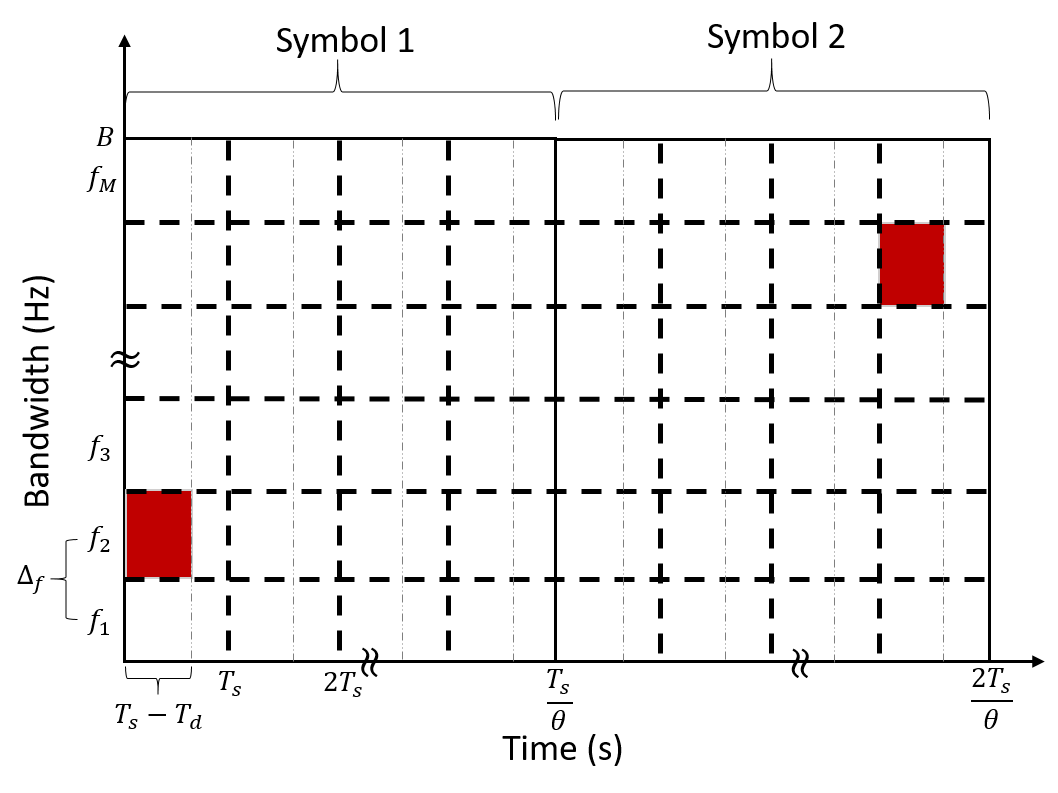}
    \caption{Grid illustrating the transmission of two WTFC symbols with 1 frequency randomly chosen from $M$ possible frequencies and duty cycle of $\theta$. Shaded regions denote transmitted frequencies. WTFC signals are not restricted to a known time period.}
    \label{fig:wtfc_grid}
\end{subfigure}
\caption{Visualization of the transmission of I-FSK and WTFC signals.}
\label{fig:ifsk_wtfc_sig}
\end{figure*}

I-FSK is a signaling scheme that combines FSK with a duty cycle, where both the transmitter and receiver have knowledge of which time period the signal is transmitted in \cite{teletar_pfsk}. 
Multiple frequencies can be transmitted using I-FSK.
WTFC combines FSK with a duty cycle, and incorporates PPM into I-FSK.
The time period in which the signal is transmitted is unknown to the receiver \cite{wtfc}.
Using WTFC, only a single frequency is transmitted.

Fig. \ref{fig:ifsk_wtfc_sig} shows a visualization of the transmission of I-FSK and WTFC signals using a grid layout with the y-axis representing frequencies, and the x-axis representing time. 
Both signaling schemes have similar setups, where there are $M$ frequencies denoted as $f_i$ with $i\in [1,M]$, a duty cycle $\theta$ with signal transmission occurring $1/\theta$ of the time, and a symbol time of $T_s$.
The separation between adjacent frequencies  is denoted as $\Delta_f = f_{i+1} - f_{i}$.
Taking into account the channel delay spread, $T_d$, the frequency separation is calculated as $\Delta_f = 1/(T_s-T_d)$ in order to maintain orthogonality between frequency tones \cite{multitone_pfsk}.

The bandwidth $B$ of the systems can be calculated from the frequency separation and the number of possible frequency tones $M$ via $B=M/(T_s-T_d)$.
After demodulation to baseband and assuming a bandwidth of $[-B/2, B/2]$, the frequencies are
\begin{equation}
\begin{split}
    f_i = -\frac{B}{2} + \frac{2i-1}{2(T_s-T_d)} =\frac{2i-M-1}{2(T_s-T_d)}.
\end{split}
\end{equation}
where $i\in[1,M]$.

\subsubsection{I-FSK Signal Model}

Fig. \ref{fig:PFSK_grid} illustrates the transmission of I-FSK signals comprised of two frequency tones with a duty cycle of $\theta$.
The transmission of the I-FSK signal is fixed to the first column of the grid, and each transmission occurs $1/\theta$ of the time.
Define $Q$ to be the number of frequencies transmitted simultaneously. 
This gives a set of possible symbols $S$ with size $|S| = \binom{M}{Q}$ as each frequency is chosen without replacement.

The transmission of an I-FSK signal over a cycle time of $[0,T_s/\theta)$ is \cite{multitone_pfsk}
\begin{equation}
    x(t) = \begin{cases}
    \sum_{l\in S_m} \sqrt{\frac{P}{Q\theta}}\exp(j2\pi f_l t),& 0\leq t\leq T_s \\
    0,& \text{otherwise}.
    \end{cases}
    \label{eqn:pfsk_transmit}
\end{equation}
where $S_m$ is the symbol that is being transmitted ($S_m\in S$), $f_l$ are the frequencies being transmitted, the set $\{l\}$ contains the frequency indices associated with symbol $S_m$, $\theta$ is the duty cycle, and $P$ is the average transmit power.
The power of the transmitted signal is $P/\theta$, and the average transmit power over the cycle time is $P$. 

Consider the transmission of the I-FSK signal over a multipath fading channel.
We assume that the channel has block-fading with a coherence time of $T_c$ and a constant delay spread $T_d$.
We also assume that the channel is underspread, $T_d \ll T_c$, which results in the channel output \cite{multitone_pfsk}
\begin{align}
    y(t) = \begin{cases}\sum_{k\in S_m}\alpha_k \sqrt{\frac{P}{Q\theta}}\exp(j2\pi f_k t) + z(t),& 0\leq t \leq T_s \\
    z(t),& \mathrm{otherwise}.
    \end{cases}
    \label{eqn:ifsk_channel_output}
\end{align}
Here, $\alpha_k$ are identical and independent zero-mean complex Gaussian random variables with $\mathbb{E}[|\alpha|^2] = 1$, and $z(t)$ is complex AWGN with power spectral density $N_0/2$ \cite{proakis}. 

\subsubsection{WTFC Signal Model}

Fig. \ref{fig:wtfc_grid} illustrates the transmission of two WTFC signals with a duty cycle of $\theta$. 
It can be seen that the time duration in which a WTFC signal is transmitted is not fixed, unlike the transmission of an I-FSK signal.

From Fig. \ref{fig:wtfc_grid}, the WTFC signal is transmitted for a duration of $T_s-T_d$ seconds, while in Fig. \ref{fig:PFSK_grid}, the I-FSK signal is transmitted for $T_s$ seconds. 
The delay spread, $T_d$, is used as a guard time and prevents two signals in adjacent time periods from overlapping at the receiver.
In scenarios in which the delay spread is unknown, a different value can be used as a guard time.
Here, we fix the delay spread to be the guard time.

\begin{figure*}[t]
\centering
\begin{subfigure}{0.46\textwidth}
    \centering
    \includegraphics[width=0.965\textwidth]{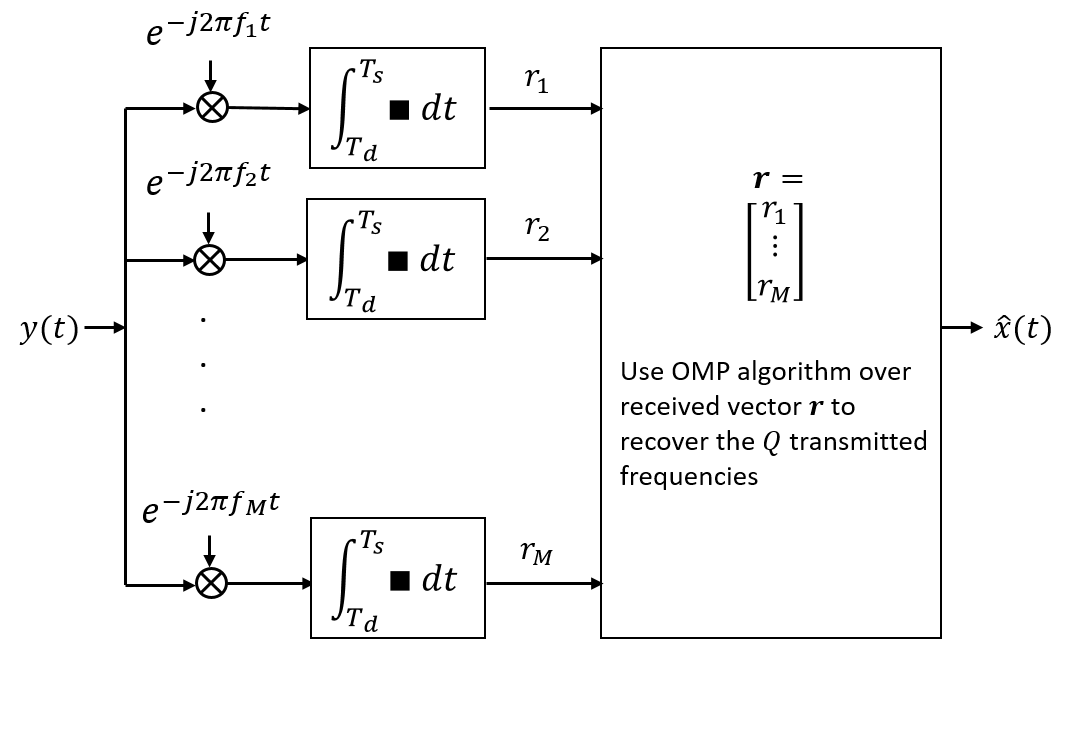}
    \caption{Matched filter receiver for I-FSK signals.}
    \label{fig:PFSK_mf}
\end{subfigure}
\hfill
\begin{subfigure}{0.49\textwidth}
    \centering
    \includegraphics[width=\textwidth]{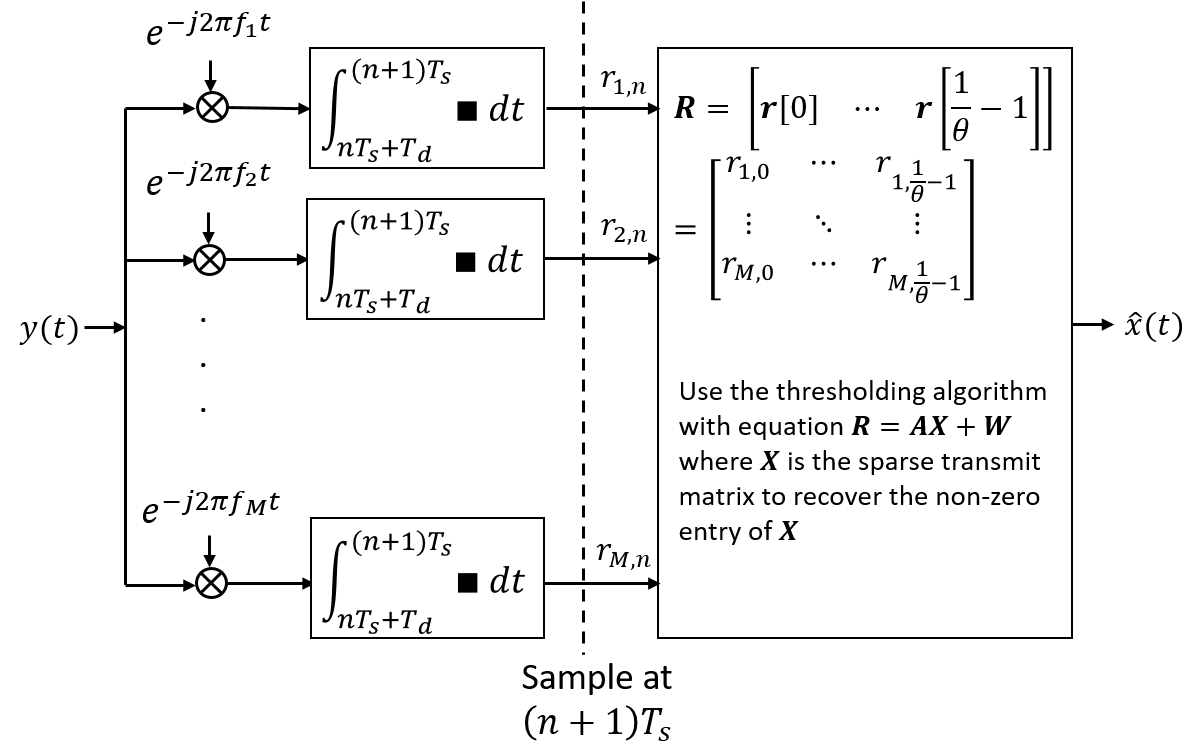}
    \caption{Matched filter receiver for WTFC signals.}
    \label{fig:WTFC_mf}
\end{subfigure}
\caption{Matched filter receivers for I-FSK and WTFC signals.}
\label{fig:mf_receiver}
\end{figure*}

The transmission of a WTFC signal over a cycle time of $[0, T_s/\theta)$ is \cite{wtfc}
\begin{align}
x_{l,k}(t) &= \begin{cases}
A\mathrm{exp}(j2\pi f_lt),& kT_s\leq t \leq (k+1)T_s-T_d \\
0 ,& \mathrm{otherwise}.
\end{cases} 
\label{eqn:wtfc_sig}
\end{align}
where the amplitude of the signal is
\begin{align}
    A &= \sqrt{\frac{PT_s}{\theta (T_s-T_d)}}.
    \label{eqn:wtfc_amp}
\end{align}
Here, $A$ is the amplitude of the transmitted signal, $f_l$ is the frequency of the tone being transmitted, $k$ is the index for the time period in which the tone is transmitted with $k\in [0,1/\theta -1]$, and $P$ is the average transmit power.

Assuming the same multipath fading channel described in the previous section, the channel output is expressed as \cite{wtfc}
\begin{equation}
    \resizebox{.44\textwidth}{!}{$\displaystyle y(t) = \begin{cases}\alpha A\mathrm{exp}(j2\pi f_lt) + z(t),& kT_s\leq t \leq (k+1)T_s \\
    z(t),& \mathrm{otherwise}.
    \end{cases} $}
    \label{eqn:wtfc_channel_output}
\end{equation}
where $\alpha$ is a zero-mean complex Gaussian random variable with $\mathbb{E}[|\alpha|^2] = 1$ \cite{proakis}, and $z(t)$ is complex AWGN with power spectral density $N_0/2$.
\section{Receivers}
\label{sec:receivers}

\subsection{Bank of Frequency-selective Filters}

Frequency-selective filters are matched filters for FSK-based signals and are used for noncoherent detection.
Frequency-selective filters perform an inner product between the received signal and the complex conjugate of the basis functions, as shown in Fig. \ref{fig:mf_receiver} \cite{proakis}.
For I-FSK, the frequency-selective filters are sampled once per duty cycle, which is shown in Fig. \ref{fig:PFSK_mf}, while for WTFC, the frequency-selective filters are sampled $1/\theta$ times per duty cycle for each time period, which is shown in Fig. \ref{fig:WTFC_mf}.

In the following subsections, we will see that the sensing equations of I-FSK and WTFC signals are almost identical.
The primary difference is that WTFC's sensing equation incorporates an additional time index.
\subsubsection{I-FSK Outputs}
The output of the $i$-th frequency-selective filter is
\begin{align}
    r_i 
    %&= \int_{T_d}^{T_s}y(t)\exp(-j 2 \pi f_i t) \text{d}t \\
    &= \int_{T_d}^{T_s}\left(\sum_{k\in S_m} \alpha_k \sqrt{\frac{P}{Q\theta}}\exp(j2\pi f_k t) + z(t)\right)\times \nonumber\\
    &\;\;\;\;\;\;\;\;\;\;\;\;\;\;\exp(-j 2 \pi f_i t) \text{d}t \nonumber \\
    %&= \int_{T_d}^{T_s}\sum_{k\in S_m} \alpha_k \sqrt{\frac{P}{Q\theta}}\exp(j2\pi (f_k-f_i) t)\text{d}t + \int_{T_d}^{T_s}w(t)\exp(-j 2 \pi f_i t) \text{d}t \\
    &= \begin{cases}
        z_i, & \text{for} \; i\notin S_m \\
        \alpha_i (T_s-T_d) \sqrt{\frac{P}{Q\theta}} + z_i, &\text{for} \; i \in S_m \\
    \end{cases}
\end{align}
where $z_i$ is a complex Gaussian random variable with zero mean and variance $N_0(T_s-T_d)$. 
%From Eq. 2.2 to 2.3, we use the condition that the frequencies are orthogonal to split between the cases where $i \in S_m$ and $i \notin S_m$.

The outputs of the frequency-selective filters, an $M \times 1$ vector $\mathbf{r}$ with entries $r_i$, can be expressed in terms of a linear equation with a sensing matrix, input vector, and noise vector.
Define $\mathbf{x}$ to be an $M \times 1$ vector with ${x}_k = \alpha_k \sqrt{\frac{P}{Q\theta}}$ for $k\in S_m$, and  ${x}_{k'} = 0$ for $k'\notin S_m$, and $\mathbf{z}$ to be a $M \times 1$ vector with entries $z_i$.
Then,
\begin{align}
    \mathbf{r} = (T_s-T_d)\mathbf{x} + \mathbf{z}.
    \label{eqn:ifsk_mf_eq}
\end{align}
The noise outputs $z_i$ are independent due to the frequencies being orthogonal.
Thus, the covariance matrix of the noise vector $\mathbf{z}$ is $\text{Cov}(\mathbf{z}) = N_0(T_s-T_d)\mathbf{I}$.
\subsubsection{WTFC Outputs}
WTFC signals can be transmitted in any time period.
This results in an additional time index in comparison to I-FSK.
Given that frequency $f_k$ was transmitted in the $m$-th time period where $i,k \in[1, M]$ and $n,m \in[0, 1/\theta-1]$, the output of the $i$-th frequency selective filter at the $n$-th time period  is
\begin{align}
    r_{i,n} &= \int_{nT_s + T_d}^{(n+1)T_s} y(t) \exp \left({-j2\pi f_i t} \right) \mathrm{d}t. \nonumber\\
    &=\begin{cases}
    z_{i,n}, & (i,n) \neq (k,m) \\
    \alpha (T_s-T_d) \sqrt{\frac{PT_s}{\theta (T_s-T_d)}} + z_{i,n}, & (i,n) = (k,m)
    \end{cases}
    \label{eqn:r_m,n}
\end{align}
where $z_{i,n}$ is a complex Gaussian random variable with zero mean and variance $N_0(T_s-T_d)$.

Similar to the I-FSK outputs case, we can express the WTFC matched filter outputs in terms of a linear equation with a sensing matrix, input vector, and noise vector.
In comparison to I-FSK, a time index is added to the vectors as the transmission of a WTFC signal is not fixed to a time period. 
Define $\mathbf{x}[n]$ to be a $M \times 1$ vector with elements $x_{i,n}$ where $x_{i,n} = \alpha \sqrt{\frac{PT_s}{\theta (T_s-T_d)}}$ for $(i,n) = (k,m)$, otherwise $x_{i,n} = 0$; $\mathbf{z}[n]$ to be a $M \times 1$ vector with entries $z_{i,n}$; and $\mathbf{r}[n]$ to be a $M \times 1$ vector where the entries are the outputs of the bank of frequency-selective filters at the $n$-th sample.
Then, the sensing equation can be written as,
\begin{align}
    \mathbf{r}[n] = (T_s-T_d)\mathbf{x}[n] + \mathbf{z}[n].
    \label{eqn:wtfc_mf_eq}
\end{align}
The noise outputs in the vector $\mathbf{z}[n]$ are independent due to the frequencies being orthogonal. 
Thus, the covariance matrix of the noise vector $\mathbf{z}[n]$ is $\text{Cov}(\mathbf{z}[n]) = N_0(T_s-T_d)\mathbf{I}$.

As shown in Fig. \ref{fig:WTFC_mf}, the $\mathbf{r}[n]$ can be joined together to form a matrix $\mathbf{R} = \begin{bmatrix}\mathbf{r}[0] & \mathbf{r}[1] & ... & \mathbf{r}[1/\theta -1]\end{bmatrix}$, which is used in the OMP algorithm to recover the original signal.

\begin{figure*}[t]
\hfill
\begin{subfigure}{0.46\textwidth}
    \centering
    \includegraphics[width=0.975\textwidth]{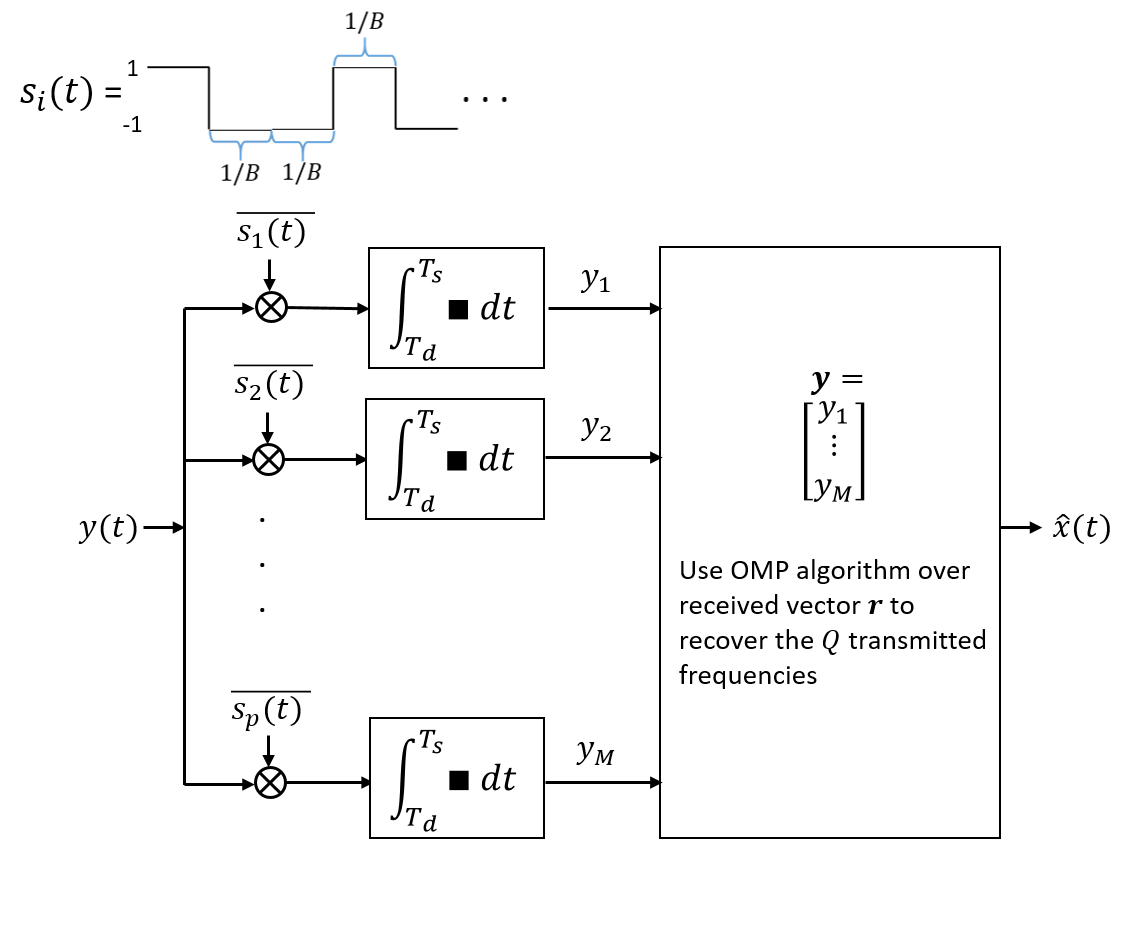}
    \caption{Compressed sensing receiver for I-FSK signals.}
    \label{fig:IFSK_cs}
\end{subfigure}
\hfill
\begin{subfigure}{0.49\textwidth}
    \centering
    \includegraphics[width=\textwidth]{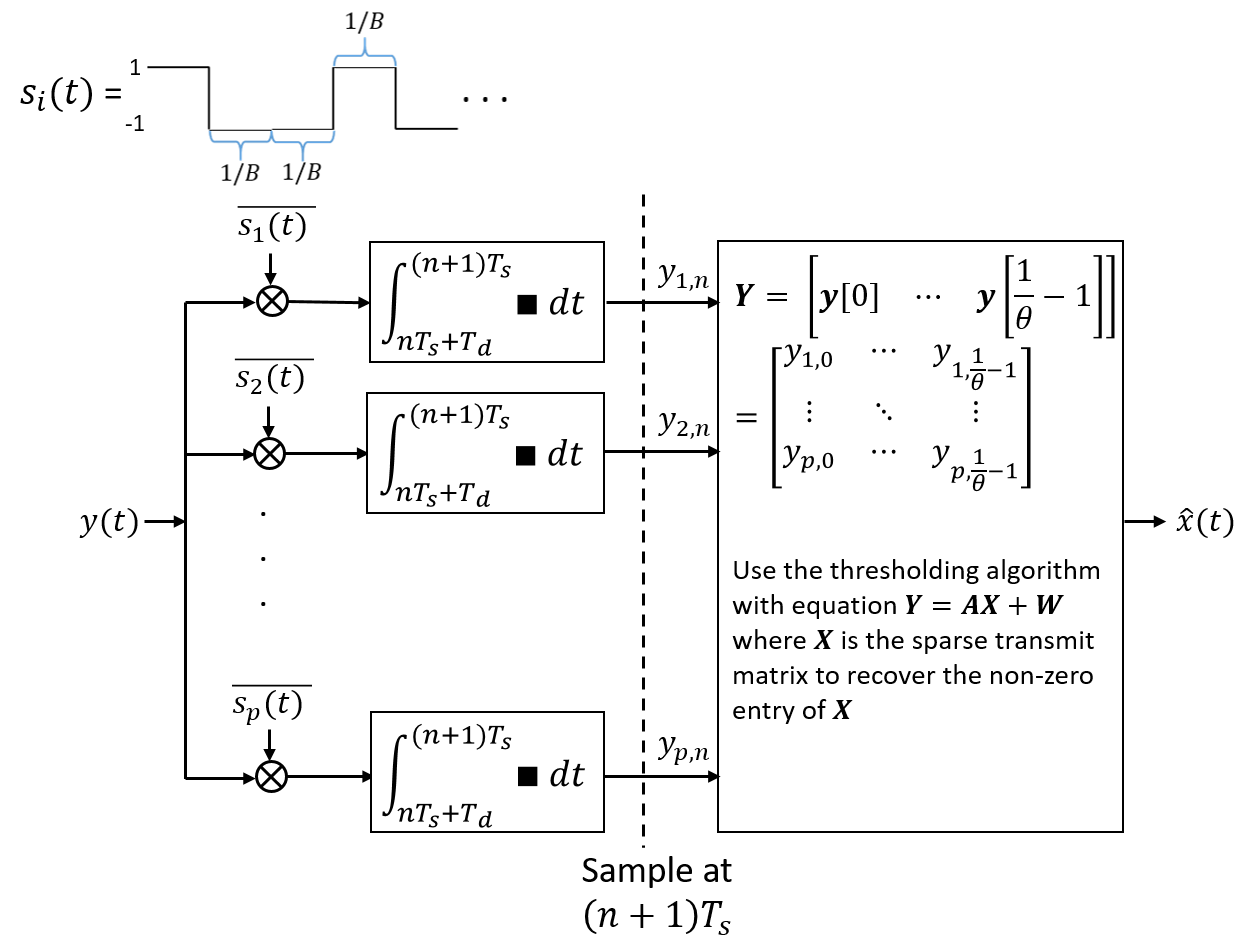}
    \caption{Compressed sensing receiver for WTFC signals.}
    \label{fig:WTFC_cs}
\end{subfigure}
\caption{Compressed sensing receivers for I-FSK and WTFC signals.}
\label{fig:cs_receiver}
\end{figure*}

\subsection{Compressed Sensing Receiver with Correlating Signals}
The sparsity of the I-FSK signals where $Q \ll M$ allows for the usage of a compressed sensing receiver that reduces the number of filters required to recover I-FSK signals.
Fig. \ref{fig:cs_receiver} shows a compressed sensing receiver with $p$ correlating signals, where $p\leq M$, for both I-FSK and WTFC.

Here, we use chipping sequences as the correlating signals.
Chipping sequences are signals composed of rectangular pulses with length less than or equal to the Nyquist rate \cite{random_demodulator, mwc}.
Fig. \ref{fig:cs_receiver} shows an example of a chipping sequence where the bandwidth is $B$ above the receivers for I-FSK and WTFC.
The chipping sequences are defined as
\begin{align}
    s_{i}(t) = \sum_{l=1}^{M} v_{il}\, \text{rect}\left(t-T_d-\frac{l-1}{B}\right),
    \label{eq:chip_seq}
\end{align} 
where $s_i(t)$ is the i-th chipping sequence with $i\in[1,p]$, $v_{il}$ are the amplitudes of the pulses and are chosen randomly with equal probability from $[-1,1]$, and rect$(t)$ is a rectangular pulse with amplitude 1 over the time duration $t=[0,1/B]$.
Each chipping sequence consists of $M$ rectangular pulses, and is non-zero over $t=[T_d,T_s]$.
In this work, once $s_i(t)$ have been randomly generated, the same $s_i(t)$ are used to recover the signals throughout the simulations.

%\begin{figure}[t] 
%\centering
%\includegraphics[width=0.7\linewidth]{Figures/chipping_seq.png}
%\caption{A chipping sequence composed of rectangular pulses of length $1/B$ seconds.}
%\label{fig:chipping_seq}
%\end{figure}

\subsubsection{I-FSK Outputs}
With the correlating signals $s_i(t)$ defined in (\ref{eq:chip_seq}), the integration of the product between the received signal and the complex conjugate of the $i$-th correlating signal is
\begin{align}
    y_i 
    %&= \int_{T_d}^{T_s}y(t) \overline{s_i(t)} \text{d}t \\
    &= \int_{T_d}^{T_s}\left(\sum_{k\in S_m} \alpha_k \sqrt{\frac{P}{Q\theta}}\exp(j2\pi f_k t) + w(t)\right)\times \nonumber \\ &\;\;\;\;\;\;\;\;\;\;\;\;\;\sum_{l=1}^{M}v_{il}\,\text{rect}\left(t-T_d-\frac{l-1}{B}\right) \text{d}t \nonumber \\
    &= \int_{T_d}^{T_s}\sum_{k\in S_m} \alpha_k \sqrt{\frac{P}{Q\theta}}\exp(j2\pi f_k t)\times
    \label{eqn:cs_total}
    \\\nonumber
    &\;\;\;\sum_{l=1}^{M}v_{il}\,\text{rect}\left(t-T_d-\frac{l-1}{B}\right) \text{d}t + \int_{T_d}^{T_s}w(t){s_i(t)} \text{d}t.
\end{align}
%\textcolor{red}{Add more steps to below equation}
The first term of (\ref{eqn:cs_total}) is
\begin{align}
    &\int_{T_d}^{T_s}\sum_{k\in S_m} \alpha_k \sqrt{\frac{P}{Q\theta}}\exp(j2\pi f_k t)\times \nonumber\\
    &\;\;\;\;\;\;\;\;\;\sum_{l=1}^{M}v_{il}\,\text{rect}\left(t-T_d-\frac{l-1}{B}\right) \text{d}t \nonumber\\
    %&= \sum_{k\in S_m}\alpha_k \sqrt{\frac{P}{Q\theta}}\sum_{l=1}^{M}\int_{T_d}^{T_s} \exp(j2\pi f_k t) \times \\\nonumber
    %&\;\;\;\;\;\;\;\;\;v_{il}\,\text{rect}\left(t-T_d-\frac{l-1}{B}\right) \text{d}t \\
    %&= \sum_{k\in S_m}\alpha_k \sqrt{\frac{P}{Q\theta}}\sum_{l=1}^{M}\sum_{g=1}^{M}\int_{T_d+(g-1)/B}^{T_d+g/B} \exp(j2\pi f_k t)v_{il}\,\text{rect}\left(t-T_d-\frac{l-1}{B}\right) \text{d}t \\
    %&= \sum_{k\in S_m}\alpha_k \sqrt{\frac{P}{Q\theta}}\sum_{l=1}^{M}\int_{T_d+(l-1)/B}^{T_d+l/B} \exp(j2\pi f_k t)v_{il}\, \text{d}t \nonumber\\
    &= \sum_{k\in S_m}\alpha_k \sqrt{\frac{P}{Q\theta}} \frac{j}{2\pi f_k} \exp\left(j2\pi f_k T_d\right)\times\\\nonumber
    &\;\;\;\;\;\left(1-\exp\left(j2\pi f_k\frac{1}{B}\right)\right) \sum_{l=1}^{M} v_{il} 
    \exp\left(j2\pi f_k \frac{l-1}{B}\right).
    \label{eqn:IFSK_first_term}
\end{align}
See App. \ref{app:cs_derivation} for the detailed derivation of this first term.

%To express the $p \times 1$ output vector $\mathbf{y}$ with entries $y_i$ in terms of a linear equation, we 
Define $\mathbf{V}$ to be a $p \times M$ matrix containing the correlating signal coefficients with $\mathbf{V}_{il} = v_{il}$, $\mathbf{D}$ to be a diagonal $M \times M$ matrix with $\mathbf{D}_{kk} = \frac{j}{2\pi f_k}\exp(j2\pi f_k T_d)\left(1-\exp\left(j2\pi f_k \frac{1}{B}\right)\right)$, $\boldsymbol{\omega}$ to be a $p \times 1$ vector with entries $\boldsymbol{\omega}_i = \int_{T_d}^{T_s}w(t)\overline{s_i(t)} \text{d}t$, and $\mathbf{F}$ to be a $M \times M$ matrix with entries of
\begin{align}
    \mathbf{F}_{lk} 
    &=  \exp\left(j2\pi f_k \frac{l-1}{B}\right) \nonumber \\
    %&=  \exp\left(j2\pi f_k \frac{l-1}{B}\right)\\
    %&= \exp\left(j2\pi \frac{2k-M-1}{2(T_s-T_d)} \frac{(T_s-T_d)(l-1)}{M}\right)\\
    &= \exp\left(j2\pi \frac{(2k-M-1)(l-1)}{2M}\right),
\end{align}
where $l,k \in [1,M]$.
It can be seen that $\mathbf{F}$ is a discrete Fourier transform matrix and therefore, $\mathbf{FF}^* =\mathbf{F}^*\mathbf{F} = M\mathbf{I}$. 
If we choose $\mathbf{V}$ as a random Bernoulli matrix, then, $\mathbb{E}[\mathbf{VV}^*] = M\mathbf{I}_{p\times p}$ and  $\mathbb{E}[\mathbf{V}^*\mathbf{V}] = p\mathbf{I}_{M\times M}$. 

Define $\mathbf{x}$ to be an $M \times 1$ vector with ${x}_k = \alpha_k \sqrt{\frac{P}{Q\theta}}$ for $k\in S_m$, and  ${x}_{k'} = 0$ for $k'\notin S_m$, and $\mathbf{w}$ to be a $M \times 1$ vector with entries $w_i$.
Then, the outputs of the compressed sensing receiver can be written as
\begin{align}
    \mathbf{y} = \mathbf{VFDx} + \boldsymbol{\omega}.
    \label{eq:cs_matrix}
\end{align}
where the noise covariance is $\text{Cov}(\boldsymbol{\omega}) = \frac{N_0}{B}\mathbf{VV}^*$ \cite{rdmud, y_xie_phd}.
Refer to App. \ref{app:cs_derivation} for the derivation.
\subsubsection{WTFC Outputs}
The WTFC compressed sensing receiver outputs have an additional time index to account for the transmission of WTFC signals in any time period.
As I-FSK and WTFC signals are both FSK-based signals, we can extrapolate from the I-FSK compressed sensing receiver outputs to find the WTFC compressed sensing outputs with an additional time index.
Recall that in WTFC, a single frequency is transmitted, which results in a single $\alpha$ and $f_k$ term.

Given that frequency $f_k$ was transmitted in the $m$-th time period for $i,k\in[1,M]$ and $n,m\in[0, 1/\theta -1]$, the output of the $i$-th correlating signal at the $n$-th time period is
\begin{align}
    y_{i, n} 
    &= \int_{T_d}^{T_s}y(t)\times \sum_{l=1}^{M}v_{il}\,\text{rect}\left(t-T_d-\frac{l-1}{B}\right) \text{d}t \nonumber\\
    &= \begin{cases}
    \int_{T_d}^{T_s} \alpha A\exp(j2\pi f_k t)
    \sum_{l=1}^{M}v_{il}\,\text{rect}\left(t-T_d-\frac{l-1}{B}\right) \text{d}t \\
    +\int_{T_d}^{T_s}w(t){s_i(t)} \text{d}t,   \;\;\;\;\;\;\;\;\text{for } (i, n) = (k,m)  \\
    \int_{T_d}^{T_s}w(t){s_i(t)} \text{d}t,  \;\;\;\;\;\;\;\;\;\;\;\,\text{otherwise}
    \end{cases}
    \label{eqn:wtfc_cs_output}
\end{align}
where $A=\sqrt{\frac{PT_s}{\theta(T_s-T_d)}}$.
%\textcolor{red}{Add more steps to below equation}
The non-noise portion of (\ref{eqn:wtfc_cs_output}) for the case $n=m$ and $i=k$ is
\begin{align}
    &\int_{T_d}^{T_s} \alpha A\exp(j2\pi f_k t) 
    \sum_{l=1}^{M}v_{il}\,\text{rect}\left(t-T_d-\frac{l-1}{B}\right) \text{d}t \nonumber\\
    &= \alpha A \frac{j}{2\pi f_k} \exp\left(j2\pi f_k T_d\right)\left(1-\exp\left(j2\pi f_k\frac{1}{B}\right)\right)\times \nonumber\\
    &\;\;\;\;\; \sum_{l=1}^{M} v_{il} 
    \exp\left(j2\pi f_k \frac{l-1}{B}\right).
    \label{eqn:wtfc_first_term}
\end{align}

We use the same notations and definitions for the matrices in the sensing equation as in the I-FSK case, as the derivations are identical with the exception of an extra time index and different signal amplitude in WTFC.
Define $\mathbf{V}$ to be a $p \times M$ matrix containing the correlating signal coefficients with $\mathbf{V}_{il} = v_{il}$, $\mathbf{D}$ to be a diagonal $M \times M$ matrix with $\mathbf{D}_{kk} = \frac{j}{2\pi f_k}\exp(j2\pi f_k T_d)\left(1-\exp\left(j2\pi f_k \frac{1}{B}\right)\right)$, $\boldsymbol{\omega}$ to be a $p \times 1$ vector with entries $\boldsymbol{\omega}_i = \int_{T_d}^{T_s}w(t)\overline{s_i(t)} \text{d}t$, and $\mathbf{F}$ to be a $M \times M$ matrix with entries $\mathbf{F}_{lk} = \exp\left(j2\pi \frac{(2k-M-1)(l-1)}{2M}\right)$.

Similar to the case of the bank of frequency-selective filters, we add an additional time index to the transmit, noise, and receive vectors.
We use the same definitions as in the case of the bank of frequency-selective filters for WTFC.
Define $\mathbf{x}[n]$ to be a $M \times 1$ vector with elements $x_{i,n}$ where $x_{i,n} = \alpha \sqrt{\frac{PT_s}{\theta (T_s-T_d)}}$ if $(i,n) = (k,m)$, otherwise $x_{i,n} = 0$; $\boldsymbol{\omega}[n]$ to be a $p \times 1$ vector with entries $\omega_{i,n}$; and $\mathbf{y}[n]$ to be a $p \times 1$ vector with the entries being the outputs of the compressed sensing receiver at the $n$-th sample.
Then, the outputs of the compressed sensing receiver can then be written as
\begin{align}
    \mathbf{y}[n] = \mathbf{VFDx}[n] + \boldsymbol{\omega}[n],
    \label{eqn:wtfc_cs_eqn}
\end{align}
where the noise covariance is $\text{Cov}(\boldsymbol{\omega}) = \frac{N_0}{B}\mathbf{VV}^*$ \cite{rdmud, y_xie_phd}.

From Fig. \ref{fig:WTFC_cs}, it can be seen that the $\mathbf{y}[n]$ are joined together to form a matrix $\mathbf{Y} = \begin{bmatrix}\mathbf{y}[0] & \mathbf{y}[1] & ... & \mathbf{y}[1/\theta-1]\end{bmatrix}$, which is used in the OMP algorithm to recover the original signal.

\subsection{Orthogonal Matching Pursuit}
A recovery algorithm needs to be used in conjunction with the derived equations to recover the original transmitted frequencies for I-FSK and WTFC.
We use the OMP algorithm, which is a greedy basis pursuit algorithm for recovering a sparse vector $\mathbf{x}$ from the equation $\mathbf{y}=\mathbf{Ax}+\mathbf{w}$ where $\mathbf{A}$ is a given sensing matrix, $\mathbf{y}$ is the given output vector, and $\mathbf{w}$ is unknown noise, and thresholding, where we multiply two matrices together and find the largest entry in the resulting matrix \cite{omp,yonina_textbook}.
OMP is used to recover I-FSK signals due to needing to recover multiple frequencies, while thresholding is used to recover WTFC signals due to needing to recover a single frequency.
%We use thresholding to recover WTFC signals, as there is only a single frequency to recover.
It should be noted that thresholding is identical to a single step OMP algorithm.
%We use OMP directly when recovering I-FSK signals due to the signals being transmitted in a fixed time period.
%In order to recover WTFC signals, instead of using inner products between two vectors, we multiply two matrices together due to the transmission of a WTFC signal being unfixed.
Below, we describe the OMP algorithm for I-FSK and thresholding for WTFC.
\subsubsection{OMP for I-FSK}
For I-FSK, we use the OMP algorithm described below.
\begin{itemize}
    \item Step 1: Initialize the signal estimator $\hat{\mathbf{x}} = \mathbf{0}$, the residual $\mathbf{r}=\mathbf{y}$, matrix $\boldsymbol{\Lambda} = \emptyset$, and indexer $i = 1$.
    \item Step 2: Find the column of $\mathbf{A}$ that maximizes $|\mathbf{A}_i^* \mathbf{r}|$, and append to matrix $\boldsymbol{\Lambda}$.
    \item Step 3: Update the residual $\mathbf{r}= (\mathbf{I} - \mathbf{P})\mathbf{y}$ where $\mathbf{P} = \boldsymbol{\Lambda}(\boldsymbol{\Lambda}^*\boldsymbol{\Lambda})^{-1}\boldsymbol{\Lambda}^{*}$.
    \item Step 4: Update index $i = i+1$, and stop when $i > Q$.
\end{itemize}
The indices associated with the columns of $\mathbf{A}$ that maximize $|\mathbf{A}_i^* \mathbf{r}|$ through the $Q$ iterations correspond to the detected frequencies.
As an example, for I-FSK with two transmitted frequencies, the OMP algorithm detects that the 2nd and $M$th frequencies were the transmitted frequencies if the 2nd and the $M$th columns of $\mathbf{A}$ maximize $|\mathbf{A}_i^* \mathbf{r}|$.
%In the case of the frequency-selective filters, the sensing matrix is $\mathbf{A} = \mathbf{I}$ and the vector being recovered is $(T_s-T_d)\mathbf{x}$.
For the compressed sensing receiver, the sensing matrix is $\mathbf{A} = \mathbf{VF}$ and the vector being recovered is $\mathbf{Dx}$.

\subsubsection{Thresholding for WTFC}

In the previous WTFC sections, output matrices $\mathbf{R}$ and $\mathbf{Y}$ were formed by combining the $1/\theta$ vector outputs of the bank of frequency-selective filters $(\mathbf{r}[n])$ and compressed sensing receiver $(\mathbf{y}[n])$, respectively. 
We can also combine the $\mathbf{x}[n]$ to form the input matrix $\mathbf{X} = \begin{bmatrix}\mathbf{x}[0] & \mathbf{x}[1] & ... & \mathbf{x}[1/\theta - 1]\end{bmatrix}$.
Note that $\mathbf{X}$ is a sparse matrix that contains a single non-zero entry.
%Due to the transmission of a WTFC signal not being fixed to a time period, the OMP algorithm needs to operate over all receiver outputs.
We use these matrices in the equation $\mathbf{Y} = \mathbf{AX} + \mathbf{W}$ where $\mathbf{A}$ is the given sensing matrix, $\mathbf{Y}$ is the given output matrix, $\mathbf{X}$ is the unknown input matrix, and $\mathbf{W}$ is the unknown noise matrix.
We use thresholding to determine the non-zero entry of $\mathbf{X}$.
%Only a single iteration is required for the algorithm as WTFC signals contain a single frequency, which greatly simplifies the required steps.

\begin{itemize}
    \item Step 1: Find the largest entry in $|\mathbf{A}^*\mathbf{Y}|$.
    \item Step 2: The row index of the largest entry corresponds to the transmitted frequency, and the column index corresponds to the transmission time period.
%    \item Step 1: Initialize the signal estimator $\hat{\mathbf{x}} = \mathbf{0}$, the residual $\mathbf{r}=\mathbf{y}$, matrix $\boldsymbol{\Lambda} = \emptyset$, and indexer $i = 1$.
%    \item Step 2: Find the column of $\mathbf{A}$ that maximizes $|\mathbf{A}_i^* \mathbf{r}|$, and append to matrix $\boldsymbol{\Lambda}$.
%    \item Step 3: Update the residual $\mathbf{r}= (\mathbf{I} - \mathbf{P})\mathbf{y}$ where $\mathbf{P} = \boldsymbol{\Lambda}(\boldsymbol{\Lambda}^*\boldsymbol{\Lambda})^{-1}\boldsymbol{\Lambda}^{*}$.
%    \item Step 4: Update index $i = i+1$, and stop when $i > Q$.
\end{itemize}

%Due to the need to perform the OMP algorithm for $1/\theta$ times to recover a WTFC signal, further alterations need to be made to detect a signal. 
%An external loop is added to apply the OMP algorithm over all $1/\theta$ time periods, and the indices and inner product of the columns corresponding to the largest inner product with the residual need to be saved for comparison at the end of the loop.
%However, as WTFC signals only transmit a single frequency, the OMP algorithm itself does not require multiple iterations.
%This process is illustrated in Figs. \ref{fig:WTFC_mf} and  \ref{fig:WTFC_cs}, and the steps are described below.

%\begin{itemize}
%    \item Step 1: Initialize the output vector $\mathbf{y}=\mathbf{Y}_i$, the inner products list $\mathbf{ip} = []$, the columns list $\mathbf{cl} = []$, and indexer $i = 1$.
%    \item Step 2: Find the column of $\mathbf{A}$ that maximizes $|\mathbf{A}_i^* \mathbf{y}|$, and append the inner product to $\mathbf{ip}$ and the column index to $\mathbf{cl}$.
%    \item Step 3: Update index $i = i+1$, the output vector $\mathbf{y} = \mathbf{Y}_{i+1}$ and stop when $i > 1/\theta$.
%    \item Step 4: Find the maximum of all entries of $\mathbf{ip}$. The index corresponding to the maximum entry is the time period during which the signal was transmitted, and the corresponding $\mathbf{cl}$ entry is the transmitted frequency.
%\end{itemize}

\begin{figure*}[t]
\centering
\begin{subfigure}{0.49\textwidth}
    \centering
    \includegraphics[width=\textwidth]{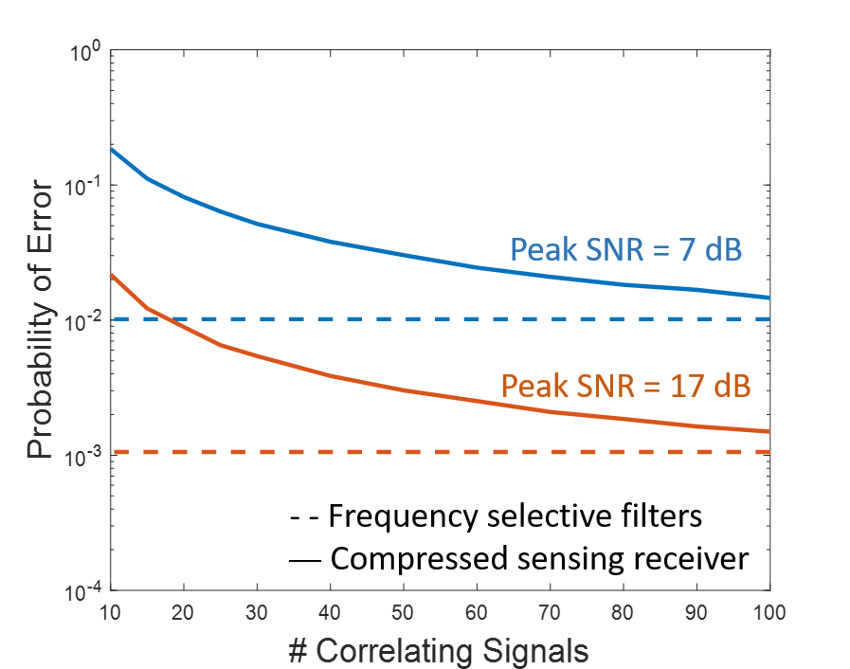}
    \caption{Recovery of an I-FSK signal with a single tone $(Q=1)$ at peak SNRs of 7 and 17 dB.}
    \label{fig:ifsk_1_tone}
\end{subfigure}
\hfill
\begin{subfigure}{0.49\textwidth}
    \centering
    \includegraphics[width=\textwidth]{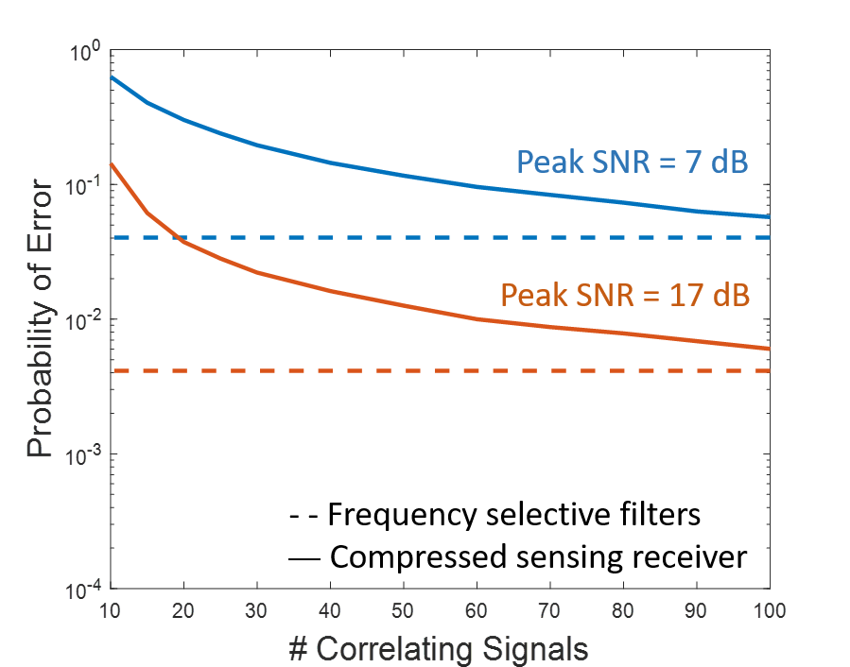}
    \caption{Recovery of an I-FSK signal with two tones $(Q=2)$ at peak SNRs of 7 and 17 dB.}
    \label{fig:ifsk_2_tone}
\end{subfigure}
\caption{Comparison between the performance of the compressed sensing receiver versus the bank of frequency-selective filters when recovering I-FSK signals with 1 or 2 tones.}
\label{fig:ifsk_tones}
\end{figure*}

For the compressed sensing receiver, the sensing matrix is $\mathbf{A} = \mathbf{VF}$, and the matrix being recovered is $\mathbf{DX}$.
As $\mathbf{D}$ is a diagonal matrix, a single entry of $\mathbf{DX}$ is non-zero, which corresponds to the non-zero entry in $\mathbf{X}$.
%As an example, 
%To clarify the last step of this algorithm, consider a toy example where $\mathbf{ip} = [0.5, 0.3, 0.8, 0.7, 0.5]$ and $\mathbf{cl} = [1, 4, 9, 3, 5]$.
%The maximum entry of $\mathbf{ip}$ is 0.8, which is the 3rd entry, and the 3rd entry of $\mathbf{cl}$ is 9. 
%The algorithm detects that the 9th frequency tone in the 3rd time period was the transmitted signal.

\section{Results and Discussion}
\label{sec:results}

\subsection{Impulsive Frequency Shift Keying}

In the simulations for the I-FSK signaling scheme, we consider the following parameters: bandwidth $B=20$MHz, noise spectral density $N_0=1$ W/Hz, average transmit power $P=10^4$ W, symbol time $T_s = 25\mu$s, delay spread $T_d = 20\mu$s, and Doppler spread $B_d = 150$Hz. 
This results in $M=100$ possible frequency tones with a frequency separation of $\Delta_f = 200$kHz.
The peak SNR was calculated in decibels using $ 10\log_{10}(P/(N_0 B \theta))$, where the fading of the channel was not taken into consideration due to its random effects.

Due to the randomness of the $\mathbf{V}$ matrix, some matrices may perform better than others.
In order to choose a sensing matrix that performs well, $10^5$ random Bernoulli matrices were generated.
The coherence associated with the matrix $\mathbf{VF}$ was calculated, and the random Bernoulli matrix that resulted in the smallest coherence was chosen as the matrix $\mathbf{V}$.

Figs. \ref{fig:ifsk_1_tone} and \ref{fig:ifsk_2_tone} show a comparison between the performance of a compressed sensing receiver with various numbers of correlating signals versus a bank of frequency-selective filters at peak SNRs of 7 and 17 dB.
Fig. \ref{fig:ifsk_1_tone} illustrates the performance of the compressed sensing receiver and the bank of frequency-selective filters when recovering 1 frequency from 100 possible frequencies, and Fig. \ref{fig:ifsk_2_tone} illustrates the performance of the receivers when recovering 2 frequencies from 100 possible frequencies.
It can be seen that increasing the number of chipping sequences for the compressed sensing receiver reduces the probability of error.
The performance of the compressed sensing receiver approaches that of the bank of frequency-selective filters as the number of chipping sequences approaches the number of frequency-selective filters.
For peak SNRs of 7 and 17 dB and the recovery of 1 or 2 tones, the performance of the bank of frequency-selective filters is $\sim$1.4-1.5 times better or $\sim$1.6 dB better than the compressed sensing receiver when the number of chipping sequences is equivalent to the number of frequency-selective filters $(p=M)$.

The difference in the performance of the compressed sensing receiver and the bank of frequency-selective filters can be attributed to the matrix $\mathbf{D}$ in the compressed sensing receiver equation.
Recall that the entries of $\mathbf{D}$ are $\mathbf{D}_{kk} = \frac{j}{2\pi f_k}\exp(j2\pi f_k T_d)\left(1-\exp\left(j2\pi f_k \frac{1}{B}\right)\right)$.
The magnitude of the entries of $\mathbf{D}$ are inversely proportional to the frequency $f_k$ and proportional to $|1-\exp\left(j2\pi f_k \frac{1}{B}\right)|$.
It is evident that the magnitude of the entries of $\mathbf{D}$ vary, which impacts the recovery of the compressed sensing receiver in contrast to the bank of frequency-selective filters.
If we compare $\mathbf{D}$ to $(T_s-T_d)$ using the ratio $M|\mathbf{D}|/(T_s-T_d)$, the $k$th entry of $M|\mathbf{D}|/(T_s-T_d)$ approaches 1 as $k$ approaches $M/2$.
There are values of $M|\mathbf{D}|/(T_s-T_D)$ that are less than 1, which indicates that some entries of $\mathbf{D}$ detrimentally impact the recovery of the signal.
Therefore, the frequency-selective filters outperform the compressed sensing receiver when the number of correlating signals is equivalent with the number of frequency-selective filters.

\begin{figure}[t]
    \centering
    \includegraphics[width = \columnwidth]{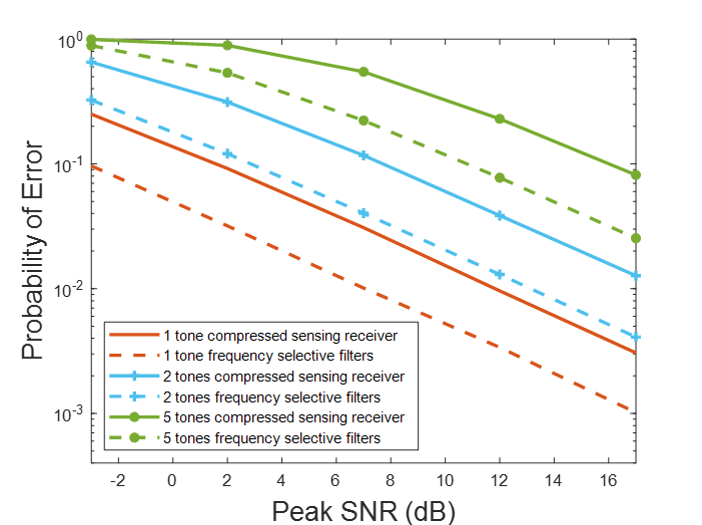}
    \caption{Comparison between the recovery of I-FSK signals with multiple tones with $p = M/2$ correlating signals for the compressed sensing receiver versus a bank of $M$ frequency selective filters.}
    \label{fig:ifsk_half_tones}
\end{figure}

Fig. \ref{fig:ifsk_half_tones} shows a comparison between the recovery of multiple frequency tones using a compressed sensing receiver with 50 correlating signals and a bank of 100 frequency-selective filters at varying peak SNRs.
The transmitted I-FSK signals contained 1, 2, or 5 frequency tones.
The results in this figure can be used to investigate how reducing the number of correlating signals impacts the performance of the compressed sensing receiver.
As the number of frequencies to recover increases, the probability of error for both the compressed sensing receiver and bank of frequency-selective filters increases.
The probability of error decreases as the peak SNR increases.
At lower peak SNRs, the performance of the bank of frequency-selective filters is not markedly better than the compressed sensing receiver due to both of their probability of errors being close to 1 for the recovery of 2 and 5 frequencies.
However, as the peak SNR increases, the ratio of the probability of error for the bank of frequency-selective filters to that of the compressed sensing receiver begins to approach a fixed range.
The ratio between the recovery of the bank of frequency-selective filters and compressed sensing receiver is consistently between 3.0-3.2 times at a peak SNR of 17 dB for the recovery of 1, 2, and 5 frequency tones.
Thus, there is approximately a 5 dB loss at larger peak SNRs when using 50 correlating signals to recovery 100 frequencies.

\subsection{Wideband Time Frequency Coding Results}
In the simulations for the WTFC signaling scheme, we explore how changing the bandwidth and the peak SNR impacts the probability of a symbol error and the recovery of the transmitted signal.
We fixed the following parameters in the simulations: delay spread $T_d = 0.3 \mu$s, Doppler bandwidth $B_d = 360$Hz, and duty cycle $\theta = 1/1000$.
The same method of choosing the $\mathbf{V}$ matrix for the recovery of WTFC signals as in the I-FSK case was used. 
We generate $10^5$ random Bernoulli matrices, and choose the random Bernoulli matrix that minimizes the coherence as the fixed matrix $\mathbf{V}$.

\begin{figure}[t]
    \centering
    \includegraphics[width = \columnwidth]{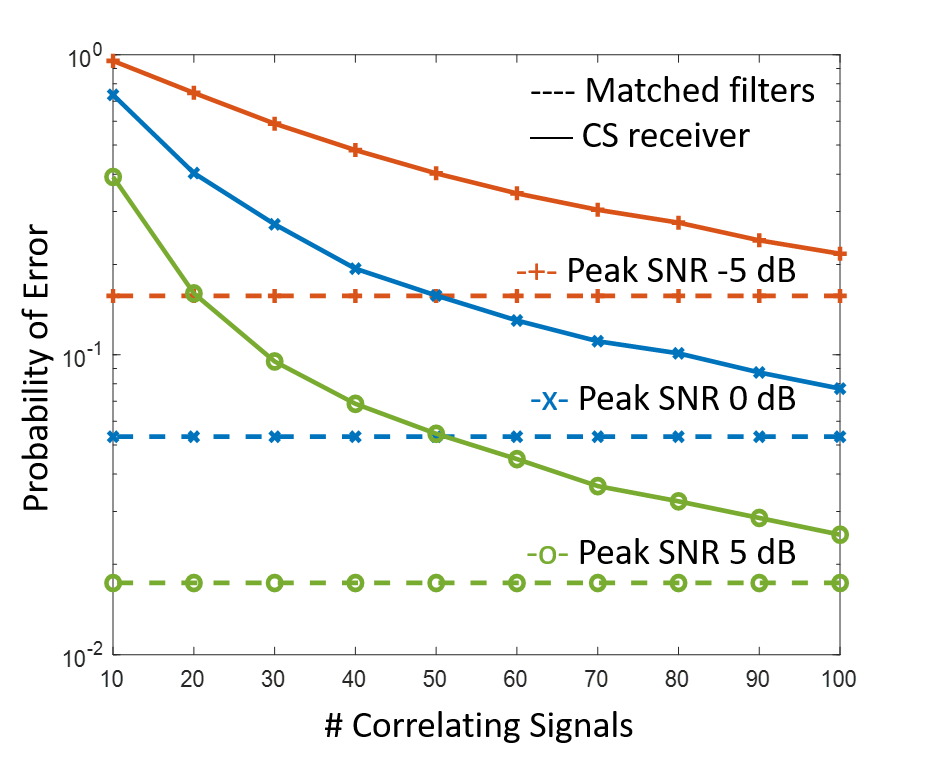}
    \caption{Comparison between the recovery of a WTFC signal using a compressed sensing receiver and a bank of frequency-selective filters at peak SNRs of $-5,0,$ and $5$dB. The following parameters were used: bandwidth $B = 400 MHz$, number of frequencies $M = 100$, and symbol time $T_s = 0.55 \mu$s.}
    \label{fig:wtfc_wideband}
\end{figure}

Fig. \ref{fig:wtfc_wideband} shows a comparison between the performance of a compressed sensing receiver versus a bank of frequency-selective filters when recovering WTFC signals at peak SNRs of $-5,0,$ and $5$ dB.
The analysis of this figure, in particular, focuses on examining the performance of WTFC and its receivers in the wideband regime, where the bandwidth of the system is $400$ MHz, and the average SNRs are $-35, -30,$ and $-25$ dB.
As the number of correlating signals used in the compressed sensing receiver increases, its performance improves.
When the number of correlating signals is equivalent with the number of frequency-selective filters $(p=M=100)$, the performance of the frequency-selective filters is $\sim$1.4-1.45 times better or $\sim$1.6 dB better than the performance of the compressed sensing receiver.
This performance matches the performance of the compressed sensing receiver when it was used to recover I-FSK signals.
The similarity between the performances is due to the sensing equations used for I-FSK and WTFC being identical, except for the recovery of a sparse transmit matrix instead of a sparse vector for WTFC (\ref{eqn:wtfc_cs_eqn}).
As the sensing equations for I-FSK and WTFC are near-identical, the difference in performance when the number of correlating signals in the compressed sensing receiver and the number of filters in the bank of frequency-selective filters are equivalent $(p=M)$ can also be explained by the $\mathbf{D}$ matrix.
The impact of the $\mathbf{D}$ matrix was discussed in the previous sub-section.
%Using the OMP algorithm to recover a sparse transmit matrix instead of a sparse vector does not seem to affect the recovery of a WTFC signal compared to a I-FSK signal.
%The additional loop that is required when recovering WTFC signals with the OMP algorithm does not seem to affect the recovery of the WTFC signal.

\begin{figure*}[t]
\centering
\begin{subfigure}{0.49\textwidth}
    \centering
    \includegraphics[width=\textwidth]{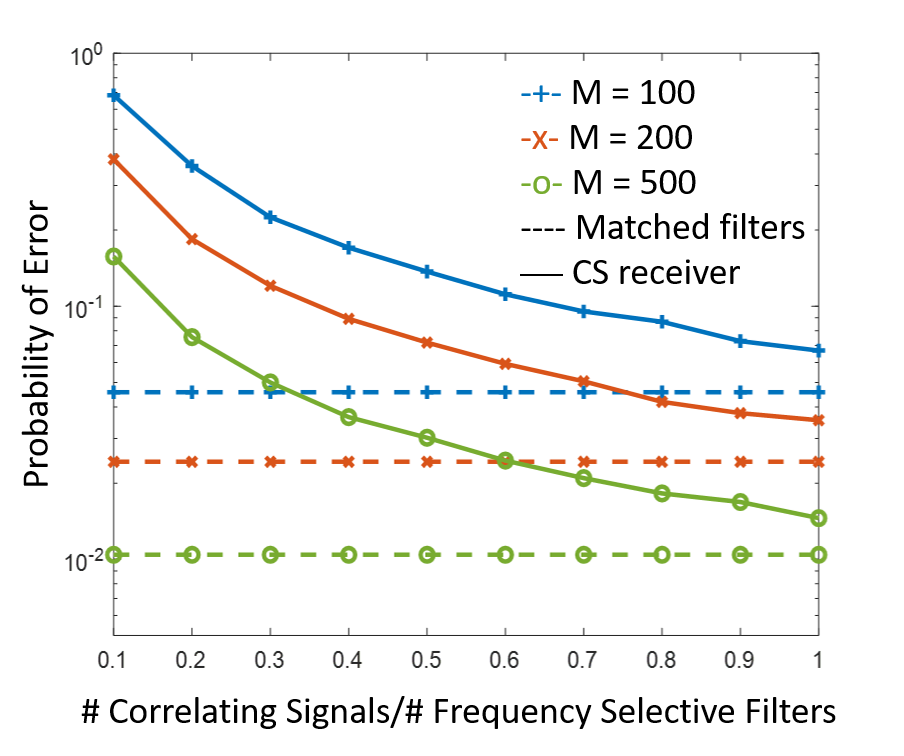}
    \caption{The probability of a symbol error associated with the matched filters and compressed sensing receiver when recovering WTFC signals.}
    \label{fig:wtfc_bandwidth}
\end{subfigure}
\hfill
\begin{subfigure}{0.49\textwidth}
    \centering
    \includegraphics[width=\textwidth]{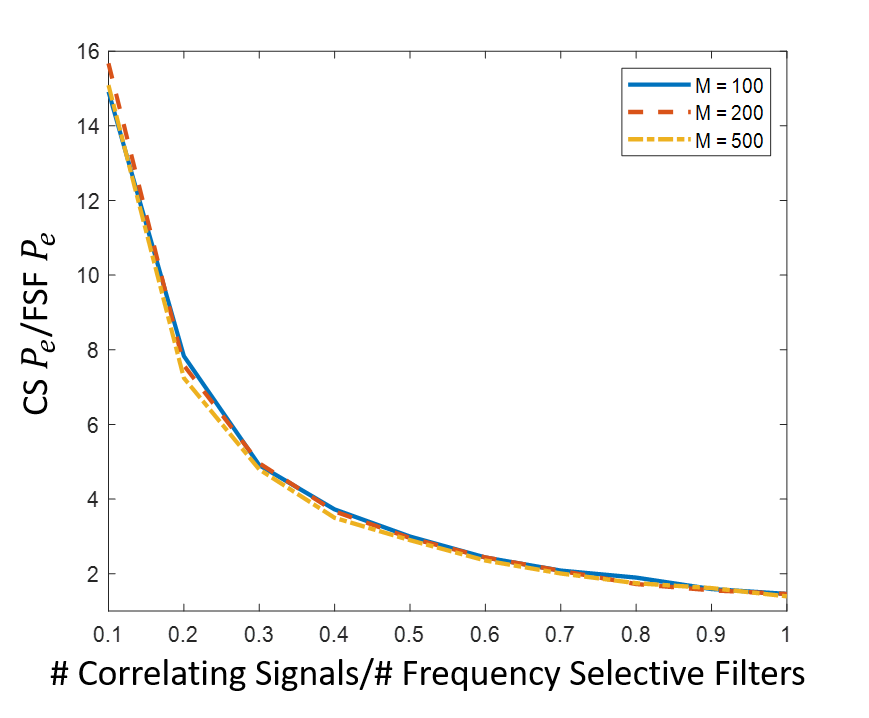}
    \caption{The ratio of the probability of symbol error of a compressed sensing receiver to that of a bank of frequency-selective filters when recovering WTFC signals.}
    \label{fig:wtfc_bandwidth_ratio}
\end{subfigure}
\caption{Comparison between the performance of the compressed sensing receiver and the bank of frequency-selective filters when recovering a WTFC signal with varying number of possible frequencies. The following parameters were used: symbol time $T_s = 5.3 \mu$s, bandwidths of $B = 20, 40, 100$MHz respectively, and peak SNR $4$ dB. This gives a separation between adjacent frequencies of $\Delta_f = 200$kHz.}
\label{fig:wtfc_M}
\end{figure*}

Fig. \ref{fig:wtfc_M} compares the performance of a compressed sensing receiver versus a bank of frequency-selective filters when recovering WTFC signals with different numbers of possible frequencies $M = 100, 200, 500$.
Fig. \ref{fig:wtfc_bandwidth} presents the probabilities of error associated with both receivers, while Fig. \ref{fig:wtfc_bandwidth_ratio} presents the ratio of the probabilities of error of the compressed sensing receiver with varying numbers of correlating signals to that of the bank of frequency-selective filters.
Due to the varying number of possible frequencies used when generating Fig. \ref{fig:wtfc_bandwidth} and \ref{fig:wtfc_bandwidth_ratio}, the x-axis is the ratio of the number of correlating signals used to the number of total frequency-selective filters. 
For example, at the ratio of 0.5, when $M=100$, 50 correlating signals were used; when $M = 200$, 100 correlating signals were used; and when $M=500$, 250 correlating signals were used. 
This allows for the results of the compressed sensing receiver and the bank of frequency-selective filters to be easily compared when viewed on the same plot. 

According to Fig. \ref{fig:wtfc_bandwidth}, the probability of a symbol error decreases as the bandwidth and the number of possible frequencies to choose from increases.
This is due to the figures being generated with a fixed peak SNR of 4 dB and average SNR of -26 dB.
With the SNRs being fixed and the increasing bandwidth of the system, the total amount of energy of the signal also increases, which reduces the probability of an error.
Similar to what has been shown previously, as the number of correlating signals increases, the performance of the compressed sensing receiver improves.

It is interesting to note that in Fig. \ref{fig:wtfc_bandwidth}, as the number of possible frequencies and the bandwidth increases, the ratio of the probability of error associated with the compressed sensing receiver to that of the bank of frequency-selective filter seems to remain constant at given ratios of the number of correlating signals to matched filters.
Such a trend is confirmed in Fig. \ref{fig:wtfc_bandwidth_ratio}.
In Fig. \ref{fig:wtfc_bandwidth_ratio}, we use the same x-axis as in Fig. \ref{fig:wtfc_bandwidth}, and the y-axis is the ratio between the compressed sensing receiver probability of error to the bank of frequency-selective filters probability of error.
We observe that ratios between the recovery of the WTFC signal using the compressed sensing receiver to that of the frequency-selective filters are close in value for all of the simulated $M$ values of $100, 200$, and $500$.
This result indicates that given a fixed peak SNR, symbol time, and frequency separation, the performance of the compressed sensing receiver compared to that of the bank of frequency-selective filters remains the same, regardless of the bandwidth of the system.
We may be able to be exploit this property when exploring the performance of a compressed sensing receiver at higher bandwidths - due to the longer simulation times required for a larger bandwidth and larger number of frequencies, it may be beneficial to extrapolate from a ratio curve generated from a smaller bandwidth to estimate the error probability.
Note that the conditions for sparsity need to be satisfied in order to ensure the validity of using a compressed sensing receiver.

\subsection{Hardware complexity and costs}
It can be seen throughout these results that there is a performance difference between using a bank of frequency-selective filters and using a compressed sensing receiver that utilizes chipping sequences.
When the number of filters and chipping sequences are equivalent, the bank of frequency-selective filters outperforms the compressed sensing receiver.
As the number of chipping sequences is decreased, the gap between the performances of the two receivers increases.

Despite this performance gap, there is a benefit in the hardware realm to using the chipping sequences instead of the frequency-selective filters. 
The compressed sensing receiver uses random binary chipping sequences, which are otherwise known as pseudo-random noise signals.
These sequences are generated by using multiple feedback shift registers \cite{pseudorandom_noise}.
The chipping sequences are all the same frequency, and the feedback shift registers can use the same oscillator output as the clock signal input.
Thus, only a single oscillator is required for the compressed sensing receiver.

In comparison, the bank of frequency-selective filters is more complex and expensive.
There are $M$ possible frequencies that can be transmitted, and each frequency-selective filter requires an oscillator to generate the associated sine wave.
While it is possible to reduce the number of required oscillators by using frequency divider or frequency multiplier circuits to change the output frequency by an integer factor, there is no guarantee that the resulting frequencies will match with the desired frequencies. 
The frequency multiplier and divider circuits would replace some oscillators, but they still introduce additional hardware.
In addition, frequency multipliers increase the phase noise \cite{freq_mult}, which would negatively impact the recovery of the FSK signals \cite{phase_noise}.

Thus, while there is a performance gap between the bank of frequency-selective filters and the compressed sensing receiver with chipping sequences, the hardware required for implementing these receivers suggests that the compressed sensing receiver would be more practical, especially as the number of possible frequencies grows large.

\section{Conclusion}
\label{sec:conclusion}
In this work, we demonstrated the feasibility of using a compressed sensing receiver with chipping sequences as correlating signals to recover I-FSK with multiple frequencies, and WTFC signals with a single frequency.
We used a bank of frequency-selective filters as a base-line for comparison.
We derived the sensing equations associated with the bank of frequency-selective filters and the compressed sensing receiver for I-FSK and WTFC, and showed that the sensing equations are identical except for an additional time index for WTFC.
We then used the OMP algorithm to recover I-FSK signals and thresholding to recover WTFC signals.
%, with the OMP algorithm for WTFC signals recovering a sparse matrix with $1/\theta$ columns due to the transmission of the signal in any possible time period in the duty cycle.

We analyzed the performance of the compressed sensing receiver under various conditions such as changing the peak SNR, number of frequencies to recover, and number of possible frequencies to recover a single frequency from.
As the number of correlating signals used to recover the transmitted signal increases, the performance of the compressed sensing receiver improves.
When the number of correlating signals is equivalent to the number of frequency-selective filters $(p=M)$, the bank of frequency-selective filters outperforms the compressed sensing receiver by $\sim$1.6 dB in all the simulations due to a diagonal matrix in the sensing matrix that attenuates the signal that is being recovered.
Reducing the number of correlating signals in the compressed sensing receiver by half when recovering 100 frequencies results in a $\sim$5 dB loss.
Given a peak SNR and frequency separation, the ratio of the performance of the compressed sensing receiver to that of the bank of frequency-selective filters remains constant even as the bandwidth of the system is changed for WTFC signals.

\section*{Acknowledgment}
We would like to thank Vedat and Assia Eyuboglu for their generous donation and support for this project.

\appendices
\allowdisplaybreaks
\section{Derivation of the non-noise component used in the compressed sensing receiver}
\label{app:cs_derivation}
Here we derive the first term of the integral for the outputs of the compressed sensing receiver, which was used in (\ref{eqn:IFSK_first_term}) and (\ref{eqn:wtfc_first_term}):
\begin{align}
    &\int_{T_d}^{T_s}\sum_{k\in S_m} \alpha_k \sqrt{\frac{P}{Q\theta}}\times \\
    &\;\;\;\;\;\;\;\;\exp(j2\pi f_k t)\sum_{l=1}^{M}v_{il}\,\text{rect}\left(t-T_d-\frac{l-1}{B}\right) \text{d}t \nonumber \\
    &= \sum_{k\in S_m}\alpha_k \sqrt{\frac{P}{Q\theta}}\times \\
    &\;\;\;\;\;\int_{T_d}^{T_s} \exp(j2\pi f_k t) \sum_{l=1}^{M}v_{il}\,\text{rect}\left(t-T_d-\frac{l-1}{B}\right) \text{d}t \nonumber \\
    &= \sum_{k\in S_m}\alpha_k \sqrt{\frac{P}{Q\theta}}\times \\
    &\;\;\;\;\;\sum_{l=1}^{M}\int_{T_d}^{T_s} \exp(j2\pi f_k t)v_{il}\,\text{rect}\left(t-T_d-\frac{l-1}{B}\right) \text{d}t \nonumber \\
    &= \sum_{k\in S_m}\alpha_k \sqrt{\frac{P}{Q\theta}}\sum_{l=1}^{M}\sum_{g=1}^{M}\times \label{eqn:rect}\\
    &\;\;\;\;\;\int_{T_d+\frac{g-1}{B}}^{T_d+\frac{g}{B}} \exp(j2\pi f_k t)v_{il}\,\text{rect}\left(t-T_d-\frac{l-1}{B}\right) \text{d}t \nonumber\\
    &= \sum_{k\in S_m}\alpha_k \sqrt{\frac{P}{Q\theta}}\sum_{l=1}^{M}\int_{T_d+\frac{l-1}{B}}^{T_d+\frac{l}{B}} \exp(j2\pi f_k t)v_{il}\, \text{d}t \label{eqn:no_rect}\\
    &= \sum_{k\in S_m}\alpha_k \sqrt{\frac{P}{Q\theta}}\sum_{l=1}^{M} v_{il} \frac{1}{j2\pi f_k} \exp(j2\pi f_k t)|_{T_d+\frac{l-1}{B}}^{T_d+\frac{l}{B}} \\
%    &= \sum_{k\in S_m}\alpha_k \sqrt{\frac{P}{Q\theta}}\sum_{l=1}^{M} v_{il} \frac{1}{j2\pi f_k} 
%    \\
%    &\;\;\;\;\;\left(\exp\left(j2\pi f_k \left(T_d+\frac{l}{B}\right)\right)-\exp\left(j2\pi f_k \left(T_d+\frac{l-1}{B}\right)\right)\right) \\
    &= \sum_{k\in S_m}\alpha_k \sqrt{\frac{P}{Q\theta}}\sum_{l=1}^{M} v_{il} \frac{1}{j2\pi f_k} \times \\
    &\;\;\;\;\;\exp\left(j2\pi f_k \left(T_d +\frac{l-1}{B}\right)\right)
    \left(\exp\left(j2\pi f_k\frac{1}{B}\right)-1\right) \nonumber\\
    &= \sum_{k\in S_m}\alpha_k \sqrt{\frac{P}{Q\theta}} \frac{j}{2\pi f_k} \exp\left(j2\pi f_k T_d\right) \times \\
    &\;\;\;\;\;\sum_{l=1}^{M} v_{il} 
    \exp\left(j2\pi f_k \frac{l-1}{B}\right)
    \left(1-\exp\left(j2\pi f_k\frac{1}{B}\right)\right) \nonumber\\
    &= \sum_{k\in S_m}\alpha_k \sqrt{\frac{P}{Q\theta}} \frac{j}{2\pi f_k} \exp\left(j2\pi f_k T_d\right) \times \\
    &\;\;\;\;\;\left(1-\exp\left(j2\pi f_k\frac{1}{B}\right)\right) \sum_{l=1}^{M} v_{il}
    \exp\left(j2\pi f_k \frac{l-1}{B}\right). \nonumber
\end{align}
From (\ref{eqn:rect}) to (\ref{eqn:no_rect}), we use $\text{rect}\left(t-T_d-\frac{l-1}{B}\right)$ being a rectangular pulse of length $1/B$ to simplify the summation of the integrals.
With this derivation, a linear equation can be written to generate the outputs of the compressed sensing receiver.

\section{Compressed Sensing Noise Covariance Matrix}
\label{app:cs_noise}
To derive the covariance matrix of the noise outputs of the compressed sensing receiver, we use a similar process as in \cite{y_xie_phd}. 
We derive the $(i,j)$th element of the covariance matrix:
\begin{align}
    C_{ij} &= \mathbb{E}[\omega_i \omega_j] \\
    &= \mathbb{E}\left[ \int_{T_d}^{T_s}\int_{T_d}^{T_s}s_i(t)s_j(u)w(t)w(u)\text{dtdu}\right] \\
    &= \int_{T_d}^{T_s}\int_{T_d}^{T_s}s_i(t)s_j(u)\mathbb{E}\left[w(t)w(u)\right] \text{dtdu}\\
    &= \int_{T_d}^{T_s}\int_{T_d}^{T_s}s_i(t)s_j(u)N_0 \delta(t-u) \text{dtdu}\\
    &= N_0\int_{T_d}^{T_s}s_i(t)s_j(t)  \text{dt}\\
    &= N_0\int_{T_d}^{T_s}\sum_{l=1}^{L}v_{il}\,\text{rect}\left(t-T_d -\frac{l-1}{B}\right)\times \\
    &\;\;\;\;\;\sum_{l=1}^{L}v_{jl}\,\text{rect}\left(t-T_d -\frac{l-1}{B}\right)  \text{dt} \nonumber\\
    &= N_0\int_{T_d}^{T_s}\sum_{l=1}^{L}v_{il}v_{jl}\,\text{rect}\left(t-T_d -\frac{l-1}{B}\right)  \text{dt}\\
    &=N_0\sum_{l=1}^{L}\int_{T_d+(l-1)/B}^{T_d+l/B}v_{il}v_{jl} \text{dt}\\
    &= \frac{N_0}{B} \sum_{l=1}^{L} v_{il} v_{jl}.
\end{align}
We see that the covariance matrix of the noise vector for the compressed sensing receiver is $Cov(\boldsymbol{\omega}) = \frac{N_0}{B}\mathbf{VV}^T$.
\nocite{masters_thesis}




\bibliographystyle{IEEEtran}
\bibliography{main}


%\end{IEEEbiography}


%\newpage



%\vfill


\end{document}